\documentstyle[graphicx,aps,multicol]{revtex}
\begin{document}
\draft
\title{Excitation Energies and Spins of the Yrast Superdeformed Band in
$^{191}$Hg}

\author{S.~Siem,$^{1,2}$ P.~Reiter,$^1$\protect\footnote[1]{Present address: Institut f{\"u}r Kernphysik, Universit{\"a}t zu K{\"o}ln, Germany} T.~L.~Khoo,$^1$ T.~Lauritsen,$^1$ P.-H.~Heenen,$^3$  M.~P.~Carpenter,$^1$ I.~Ahmad,$^1$ H.~Amro,$^{1,4}$\protect\footnote[5]{Present address: Wright Nuclear Structure Laboratory, Yale  University, New Haven, CT 06520} I.~J.~Calderin,$^1$\protect\footnote[2]{Present address: Florida state University, Tallahassee, FL 32306}  T.~D{\o}ssing,$^5$
T.~ Duguet, $^1$  S.~M.~Fischer,$^{1}$\protect\footnote[3]{Present address: Department of Physics, DePaul University, Chicago, IL 60614}
U.~Garg,$^6$ D.~Gassmann,$^1$\protect\footnote[4]{Present address: IBM, Germany} G.~Hackman,$^1$\protect\footnote[8]{Present address: TRIUMF, Vancouver, BC, Canada} F.~Hannachi,$^7$\protect\footnote[6]{Present address: CEN Bordeaux-Gradignan, IN2P3-CNRS, F-33175 Gradignan Cedex, France} K.~Hauschild,$^7$ R.~V.~F.~Janssens,$^1$ B.~Kharraja,$^6$\protect\footnote[7]{Present address: 159 Main Street, Stoneham, MA 02180, USA}  A.~Korichi,$^7$  I-Y.~Lee,$^8$  A.~Lopez-Martens,$^7$ A.~O.~Macchiavelli,$^8$ E.~F.~Moore,$^{1,4}$ D.~Nisius,$^1$\protect\footnote[9]{Present address: Bio-Imaging Research, Inc. Lincolnshire, IL 60069} C.~Sch{\"u}ck $^7$\\
$^1$Argonne National Laboratory, Argonne, Illinois 60439, USA \\ 
$^2$Department of Physics, University of Oslo, N-0316 Oslo, Norway \\  
$^3$Service de Physique Nucleaire Theorique, B-105020 Brussels, Belgium and Oak Ridge National Laboratory, Oak Ridge, Tennessee 37831, USA\\
$^4$North Carolina State University, Raleigh, North Carolina 27695, USA \\
$^5$Niels Bohr Institute, DK-2100 Copenhagen, Denmark, \\
$^6$University of Notre Dame, Notre Dame, Indiana 46556, USA \\
$^7$C.S.N.S.M, IN2P3-CNRS, bat 104-108, 91405 Orsay Campus, France \\ 
$^8$Lawrence Berkeley National Laboratory, Berkeley, California 94720, USA}
\date{\today}
\maketitle

\begin{abstract}
The excitation energies and spins of the levels in the yrast superdeformed band of $^{191}$Hg have been determined from two single-step $\gamma$ transitions and the quasi-continuum spectrum connecting the superdeformed and normal-deformed states. The results are compared with those from theoretical mean-field calculations with different interactions. 
A discussion 
 of pairing in superdeformed states is also included.
\end{abstract}


\begin{multicols}{2}

\section{Introduction}
A comprehensive understanding of superdeformed~(SD) bands requires knowledge of the quantum numbers (spin and parity) and excitation energies of the levels in the second well.  In particular, these quantities allow for stringent tests of 
configuration assignments and, more importantly, of the ability of theory to calculate shell-correction energies at large deformation. However, although more than 250 SD bands are known in the A=150 and 190 regions\cite{jans91,18}, only a few SD bands in $^{194}$Hg\cite{torb94,hac97}, $^{194}$Pb\cite{lope94,boss97}, $^{192}$Pb\cite{nabb97,wilson03} and in 
$^{152}$Dy\cite{torb02,torb02a}, have the spins and excitation energies determined through one-step linking transitions.

The yrast SD band in $^{191}$Hg was the first one to be discovered in the $A~=~190$ region~\cite{moor90}. It is especially interesting to obtain the spins and excitation energies for an odd-A SD band which, combined with data on neighboring even-even nuclei, can give information on the relative pair correlation energies in normal-deformed (ND) and superdeformed (SD) states. So far, the main information on single-particle configurations has come from detailed analyses of the $J^{(2)}$ dynamic moments of inertia of the SD bands. 
With a knowledge of the level energies and the associated quantum numbers, calculations can be put to more extensive tests and information can be obtained on properties such as particle alignment.

The work of Vigezzi \emph{et al.}\cite{Vige90} and recent improvements by Refs.~\cite{weider,barret} laid the theoretical foundation for treating the coupling of an isolated, cold SD state with a high-density of hot compound ND states,
which forms the basis of the decay mechanism. The decay of SD bands happens suddenly, typically out 
of  one to two SD states in the mass 190 region. 
One possible mechanism responsible for this sudden decay out of the SD band, proposed in Ref.~\cite{aberg99}, is 
chaos-assisted tunneling.
When the SD band decays, most of the strength is fragmented over numerous pathways, thus forming a quasi-continuum spectrum~\cite{henr94,khoo-lect} with sharp lines at high energy, which are due to direct decay to low-lying discrete ND levels. The decay spectrum from SD states is similar\cite{khoo-lect} to the spectrum following the decay of neutron-capture states\cite{lynn68}. In both cases, the decay originates from a discrete point in excitation energy and spin and  proceeds to a multitude of final states. One way of 
determining the spins and excitation energies of SD bands is to analyze the quasi-continuum decay spectrum connecting SD and ND states. 

The technique to extract the quasi-continuum decay spectrum was pioneered on $^{192}$Hg in the work of Henry \emph{et al.}\cite{henr94}. An improved method is described in detail by Lauritsen \emph{et al.}\cite{torb99} and is the one 
used in this paper. The method has been successfully tested in the case of SD band 1 in $^{194}$Hg\cite{torb99}, where it determined the same spins and excitation energies as those known from several one-step $\gamma$-ray transitions connecting the SD band to known ND states\cite{torb94,hac97}. The results from the quasi-continuum analysis are an important complement in cases where only one or two decay pathways are known. However, in most instances, 
one-step transitions are not observed, and it is then the only available option. 
 
In this work the quasi-continuum spectrum following the decay of the yrast SD band in $^{191}$Hg has been extracted. From this spectrum, we determine the excitation energy and spin of the SD band and also derive information on pairing in normal-deformed states. 
We also present two 1-step decay pathways, which directly connect the yrast SD band in $^{191}$Hg with known 
yrast levels in the ND level scheme. 
It will be shown that the results from the two methods agree very well and, thus, we can make a confident assignment of the spin and excitation energy of the yrast SD band in $^{191}$Hg. 

The experimental results are compared with theoretical calculations based on the Hartree-Fock-Bogoliubov (HFB) theory with several Skyrme interactions.
We shall also extract and discuss information on pairing in SD states, by comparing the present results with those from the even-even Hg nuclei.

\section{Experiment}
Superdeformed states in $^{191}$Hg were populated using the $^{174}$Yb($^{22}$Ne,5n)$^{191}$Hg reaction. The experiment was performed with the GAMMA\-SPHERE array\cite{lee90}, which had 96 Compton-suppressed Ge detectors at the time of the experiment. The 120~MeV $^{22}$Ne beam was provided by the 88'' Cyclotron at Lawrence Berkeley National Laboratory. The 3.1~mg/cm$^2$ $^{174}$Yb target had a 6.8~mg/cm$^2$ $^{197}$Au backing to stop the recoiling nuclei. The decay-out $\gamma$ rays were emitted after the recoils came to rest in the backing, so that the transitions will correspond to sharp lines rather than Doppler broadened ones. A total of $2$ x $10^9$ triple- or higher-fold coincidence events were collected.

\section{One-step transitions}
\begin{figure}[hbt]
\centering
\includegraphics[angle=-90,width=5in]
             {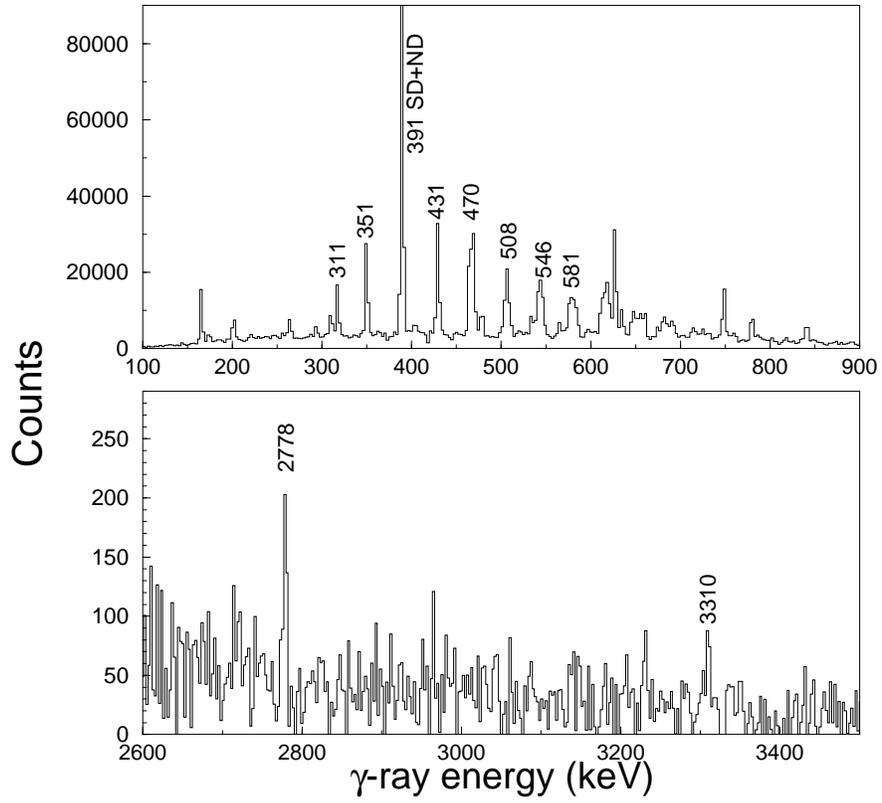}
\vspace*{1.0cm}
\caption{The $\gamma$-ray spectrum gated on clean pairs of SD transitions. The 391 keV transition is a doublet, occuring both as a SD and a ND transition. The SD transitions at the highest spins are Doppler broadened. The lower panel gives the high-energy part of the spectrum, which reveals the two 1-step decay $\gamma$ lines. The peak energies are given in keV.}
\label{fig:link}
\end{figure}

\begin{figure}[hbt]
\centering
\includegraphics[angle=-0,width=5.in] 
               {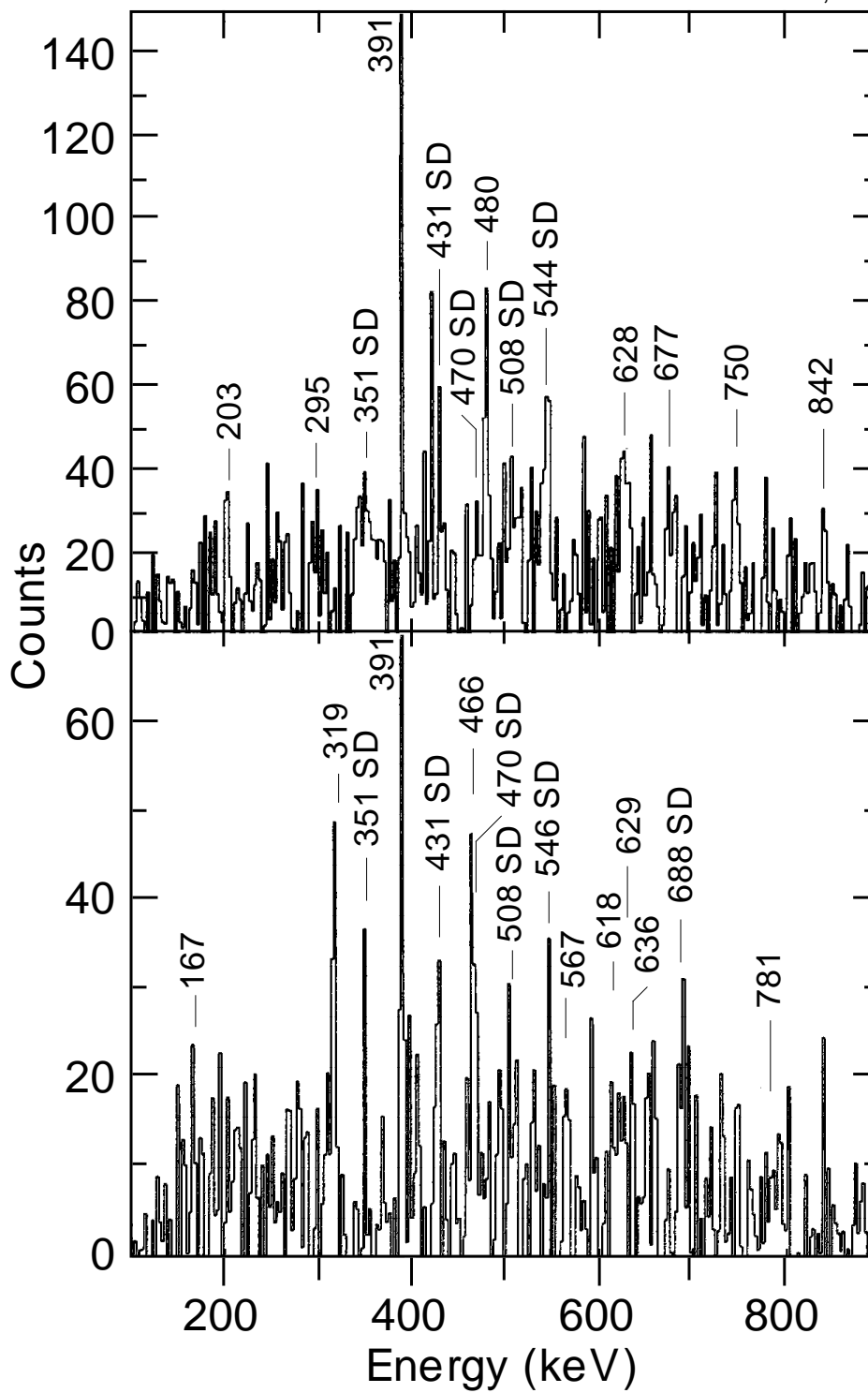}
\caption{$\gamma$-ray spectra obtained by demanding coincidences between SD transitions and the 2778 keV (top frame) or 3310 keV (bottom frame) one-step transitions. The peak energies are given in keV; peaks labeled SD belong to the SD band, while the others are ND yrast transitions expected to be seen in coincidence. The intensities of the SD transitions are distorted since  some of them are used as coincidence gates. The 391 keV transition is a doublet, occuring both as a SD and a ND transition.}
\label{fig:link-sd}
\end{figure}

\begin{figure}[htb]
  \centering
  \includegraphics[angle=0,width=5.in]
                {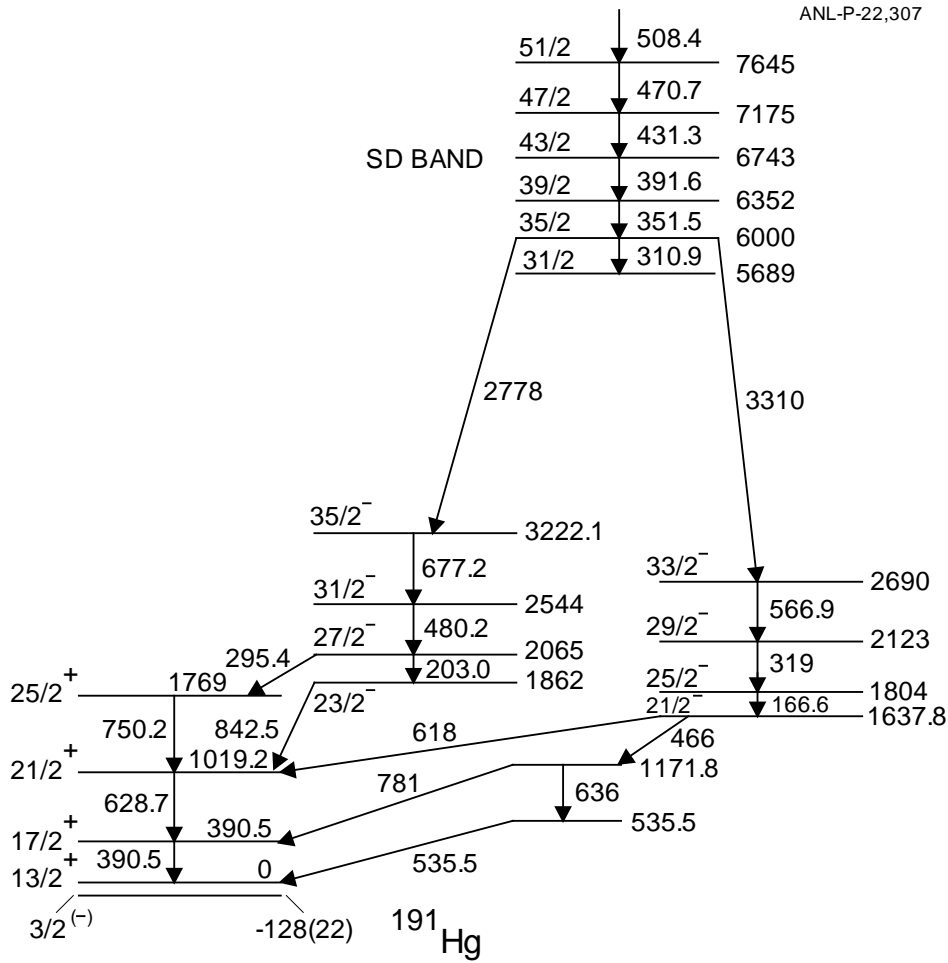}
\vspace*{1.0cm}
\caption{Partial level scheme, showing the one-step decay pathways 
connecting SD and ND levels. The intensities are 0.8\% for the 2778 keV transition or 0.4\% for the 
3310 keV transition, of the SD band intensity.
To simplify the level scheme only levels fed by the 
SD band are included and the low-intensity branches 
have been left out.  The energy of the 13/2$^+$ state, which is the termination of the SD band 
decay, has been set 
at zero since (a) it facilitates the comparison of experimental and theoretical
SD excitation energies with respect to this state 
and (b) it circumvents the uncertainty in its energy (128$\pm$22~keV, given in Ref.~\protect\cite{Schwarz,Audi}).}
\label{fig:level}
\end{figure}

The $\gamma$-ray spectrum, obtained with pairwise coincidence gates on SD transitions, is shown in Fig.~\ref{fig:link}.
The lower panel presents the high-energy part of the spectrum, in particular, the two transitions at 2778 and 3310~keV, which will be shown to connect SD and known ND states\cite{hubel,Ye}. The stronger 2778-keV transition has an area of 6$\sigma$, while the 3310-keV transition has an area of 3$\sigma$, where $\sigma$ is the statistical uncertainty. The intensities of the 2778~keV and 3310~keV lines are 
0.8 \% and 0.4 \%, respectively, of the maximum SD intra-band intensity. Figure~\ref{fig:link-sd} gives the coincidence spectra gated on a SD line and either the 2778-keV (upper figure) or 3310-keV (lower figure) transition.  Even though the statistics are low, the intensities of the ND lines suggest that the 2778-keV transition feeds the known ND 35/2$^-$ yrast level at 3222~keV.  The 3310-keV transition has been assigned to feed a 33/2$^-$ known ND level at 2690~keV.  The deduced decay scheme is shown in 
Fig.~\ref{fig:level}. 

On the basis of the coincidence data, both one-step transitions place the deexciting SD level, i.e. the one fed by the 351~keV SD transition, at 6000~keV - see Fig.~\ref{fig:level}. The angular distribution coefficient of the stronger one-step line (2778~keV), A$_2 =$~0.57$\pm$~0.48, is consistent with a $\Delta$I~=~0 dipole assignment, suggesting a 35/2~$\hbar$ spin assignment for the level fed by the 351~keV SD transition. We rule out the possibility of it being a stretched E2 transition, because that would require M3 multipolarity for the 3310-keV one-step transition. The spin is consistent with a favored $\alpha = - \frac{1}{2}$, $\em{j_{15/2}}$ configuration assignment, which is calculated to be yrast at high spin~\cite{carp95}. The experimental data do not allow for a parity assignment. However, the $\em{j_{15/2}}$ 
configuration assignment requires the SD band to have negative parity, implying M1 multipolarity for the one-step transitions. This M1 assignment is discussed later. 

The partial level scheme in Fig.~\ref{fig:level} shows the levels fed in the decay of the SD band.  
The energy of the 13/2$^+$ state is set at zero to facilitate the comparison of the experimental and 
theoretical SD excitation energies and to circumvent the 22-keV uncertainity in its excitation 
energy~\cite{Schwarz,Audi}.  Hence, the 3/2$^{(-)}$ ground state, which is not populated by the 
SD band, has an energy of -128(22)~keV in Fig.~\ref{fig:level}.

\section{Quasi-continuum analysis}
\begin{figure}[hbt]
  \centering
\vspace*{2.0cm}
  \includegraphics[angle=-90,width=7.in]
               {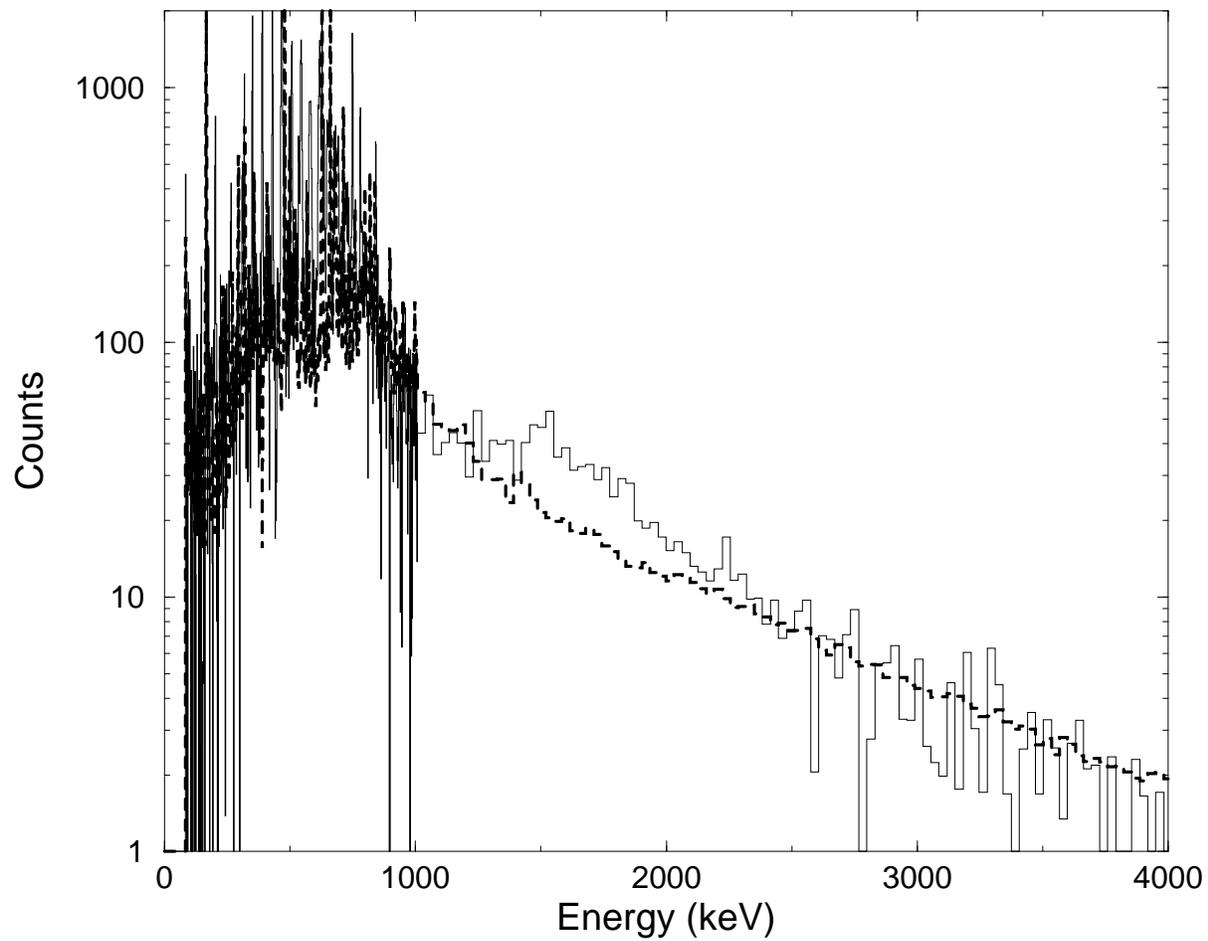}
\vspace*{1.0cm}
\caption{The total $\gamma$-ray spectrum for $^{191}$Hg, obtained from pairwise gates on SD (solid line) and ND (dotted line) transitions. 
For $E_{\gamma}$ below 1~MeV, the dispersion is 1.33~keV/channel, above 
1~MeV it is 32~keV/channel.}
\label{fig:SDND}
\end{figure}

A method has been developed at Argonne\cite{henr94,torb99} 
to isolate the quasi-continuum $\gamma$-ray spectrum connecting the SD and ND states. To confirm the results from the one-step linking transitions, this method, which is described in detail in Ref.\cite{torb99}, was followed here. First, the data were sorted with double coincidence gates on SD transitions to obtain clean spectra. Only double gates which produce clean SD spectra, without significant contamination by ND transitions, were used. The background subtraction was done using the FUL method\cite{crow95}. Corrections were carried out for $\gamma$-ray summing\cite{radford} and for neutron interactions in the detectors\cite{holz87}. The spectra were unfolded~\cite{radford} to eliminate contributions from Compton-scattered $\gamma$ rays and corrected for the detector efficiency.
The area of the spectrum was then normalized to multiplicity by requiring that the sum of the intensities of transitions feeding the ground state is 1. The 390~keV line is a doublet composed of a SD 
transition and a transition feeding the ND ground state. The 390-keV SD component is taken to have multiplicity 1, suggested by the plateau in the intensity of the SD transitions\cite{moor90,carp95}. 
The total normalized $\gamma$ spectrum is shown on a logarithmic scale in Fig.~\ref{fig:SDND}, together with the equivalent spectrum, obtained by gating on two ND yrast lines. Above 1~MeV, the spectra are contracted to 32~keV/channel and below that to 1.33~keV/channel. There is clearly extra yield in the SD gated spectrum between 1 and 2.5~MeV, which comes from the decay out of the SD band\cite{henr94,torb99}.

The discrete peaks below 800~keV are subtracted from the continuum spectrum. They can be identified as transitions either along the yrast SD band or near the ND yrast line (including previously unassigned transitions). 
The remaining continuum spectrum contains contributions from components (of statistical, quadrupole and M1/E2 nature) that feed the SD band, in addition to the sought-after decay-out spectrum. 
To extract the decay-out spectrum, the feeding components, starting with the one of statistical nature have to be subtracted. The feeding component of statistical nature cannot be disentangled experimentally from the decay-out continuum. Instead, it is obtained from a Monte Carlo simulation
 of the feeding of the SD band in $^{191}$Hg. The Monte Carlo code is described in Refs.~\cite{torb92,khoo92}. One of the inputs in the calculation of the statistical spectrum was the shape of the entry distribution. This was not measured for $^{191}$Hg in the experiment; so the shape of the entry distribution for $^{192}$Hg from Ref.~\cite{torb92} was used. The distribution was shifted to have the right average entry spin and excitation energy (33.9(1.7)~$\hbar$ and 13.8(0.7)~MeV), as found from the analysis of quasi-continuum $\gamma$ rays feeding all states in $^{191}$Hg. The average entry point for cascades feeding only the SD band was 44.1(2.2)~$\hbar$ and 17.3(0.9)~MeV, as given in Table~\ref{tab:comp}. 
 
\begin{figure}[hbt]
  \centering
\vspace*{2.0cm}
  \includegraphics[angle=-90,width=5.in]
               {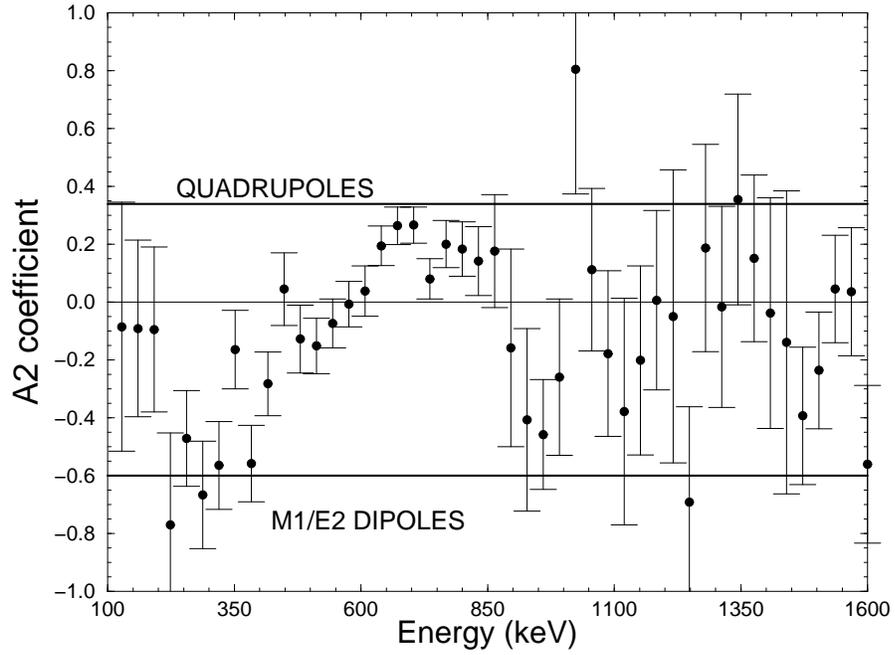}
\vspace*{1.0cm}
\caption{The $A_2$ angular distribution coefficients derived after the statistical feeding spectrum was subtracted. This figure is shown in the moving center-of-mass system, so the $A_2$ coefficients for $E_{\gamma} > 850$~keV, which are measured for $\gamma$ rays emitted after the nucleus has come to rest, are compromised.}
\label{fig:A2}
\end{figure}

\begin{figure}[hbt]
  \centering
  \includegraphics[angle=0,width=5.in]
                {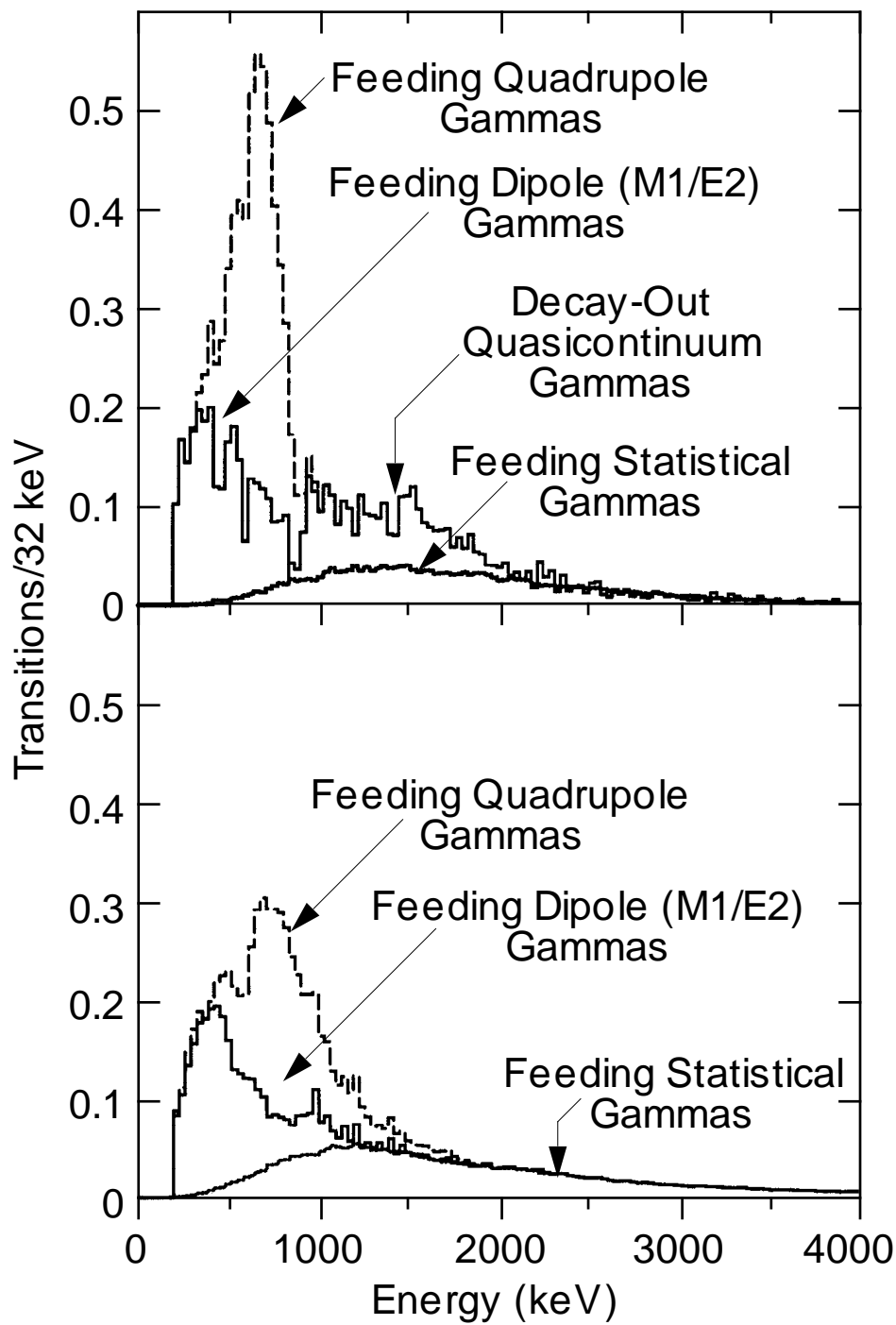}
\caption{The different components of the quasicontinuum $\gamma$ spectra in $^{191}$Hg for decays going through 
the yrast SD band (upper figure) and for decays through all (mostly ND) states (lower figure). The 
spectrum for feeding statistical transitions are from Monte Carlo simulations; all other spectra are 
from experimental data.}
\label{fig:QC}
\end{figure}

The feeding components of the spectrum are Doppler shifted, since the speed of the recoiling compound nucleus  is v/c = 0.019 (for $^{191}$Hg nuclei formed halfway through the target). To take this into account, the spectra were transformed into the  center-of-mass system.
After the statistical feeding component was removed, the quadrupole and dipole
feeding components could be separated based on the A$_2$ coefficient of the angular distribution in the center-of-mass system - see Fig.~\ref{fig:A2}. The low-energy component ($E_{\gamma}<600$~keV) is characterized by large negative $A_2$ coefficients, indicating M1/E2 nature (as seen also in $^{192,194}$Hg\cite{torb99}). After extraction and subtraction of the quadrupole E2 component, the dipole M1/E2 feeding component and decay-out component remain.
A sharp drop around 850~keV in the M1/E2 spectrum (Fig.~\ref{fig:QC}) and the drop in the A$_2$ coefficients in the same energy region indicates the presence of two components. The upper component is assigned to the decay-out of the SD band, following Refs.~\cite{henr94,torb99}.

The different components of the continuum spectrum are presented in the upper part of Fig.~\ref{fig:QC}. The energy and spin removed, on average, by the different $\gamma$-ray components are listed in Table~\ref{tab:comp}. 
For comparison, the different components of the spectrum feeding all states (mostly of ND nature) in $^{191}$Hg, from a similar analysis with two gates on yrast ND transitions (390.5 keV and 750.2 keV), is shown in the lower part of Fig.~\ref{fig:QC}. 
There are two notable differences in the two spectra in Fig. 6.  First, the     quadrupole component feeding SD states has lower energy (0.70 \emph{vs.} 0.77 MeV), is narrower and has larger multiplicity (4.0 \emph{vs.} 3.1) than that feeding ND  states.   The differences in this component, which arise from excited $\gamma$     cascades, are attributed to the larger collectivity in the SD well.  Second, the top spectrum has an additional component, starting at 0.8 and  
extending to 3.3 MeV, which arises from the decay-out quasicontinuum $\gamma$ rays connecting SD and ND states.  

\begin{figure}[htb]
  \centering
\vspace*{2.0cm}
  \includegraphics[angle=-90,width=7.in]
               {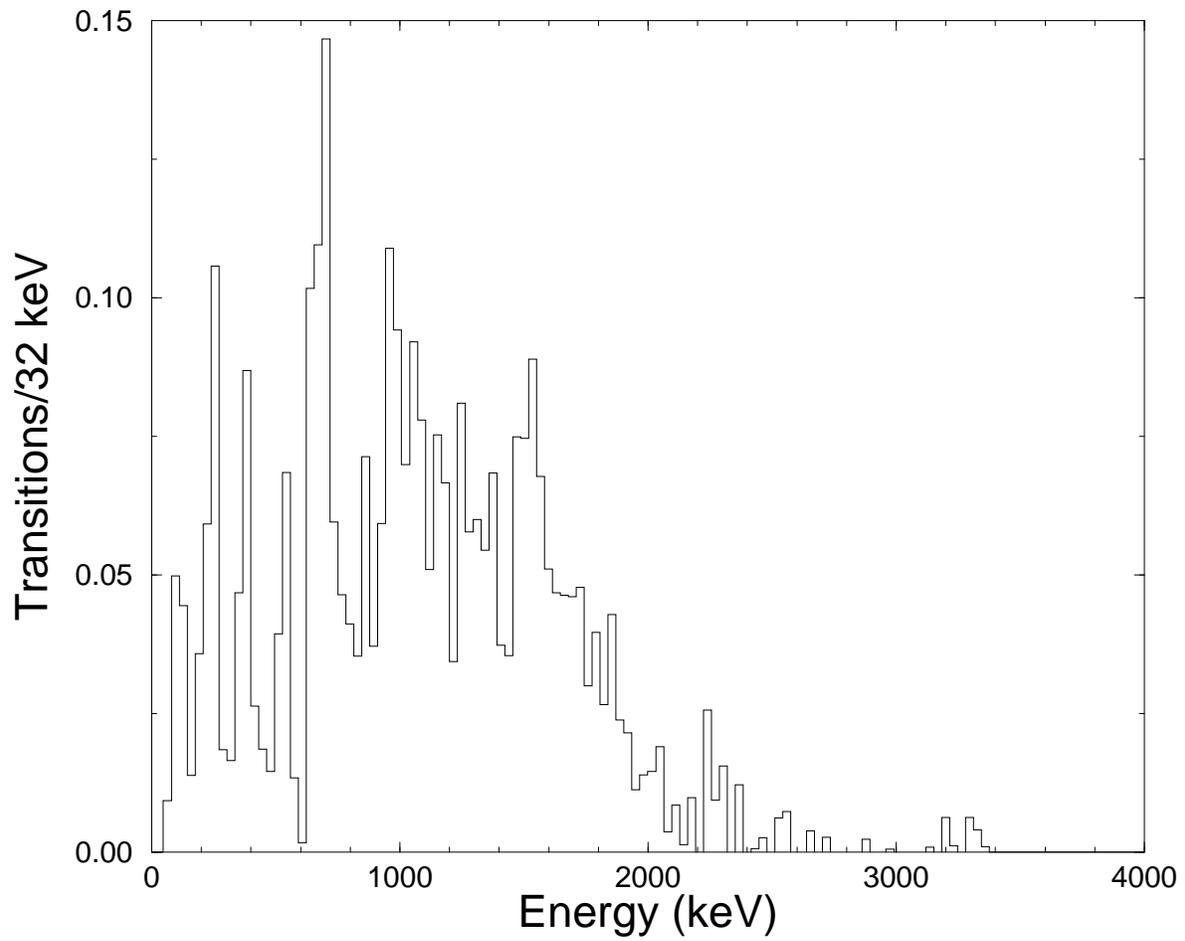} 
\vspace*{1.0cm}
 \caption{The quasicontinuum spectrum connecting the yrast superdeformed and 
 normal-deformed states. Below 800 keV the spectrum is made up of only the
non-yrast discrete lines.  Transitions along the ND yrast line, which follow the linking transitions,
 are not shown.}
\label{fig:doyr}
\end{figure}

The decay-out spectrum connecting the SD and yrast ND states (including statistical and discrete non-yrast transitions) is given in Fig.~\ref{fig:doyr}. From Monte Carlo simulations\cite{torb99,torb92} of the statistical decay, it is found that each quasi-continuum $\gamma$ ray removes 0.5(1)$\hbar$ of spin.
The energy and spin removed, on the average, by the decay-out components are found by:
\begin{equation}
\sum_{i} \Delta E_i = \langle E_{\gamma}\rangle \langle M\rangle
\end{equation} and 
\begin{equation}
\sum_{i} \Delta I_i = \langle \delta I_{\gamma}\rangle \langle M\rangle
\end{equation}
where $\langle E_{\gamma}\rangle$ and $\langle \delta I_{\gamma}\rangle$ are the average energy and spin removed per $\gamma$ ray (for a given component $i$) and $\langle M\rangle$ is the average multiplicity of this component.
The total $\gamma$-ray spectrum connecting SD and yrast ND states removes $\Delta E=3.4(2)$~MeV and $\Delta I=3.0(6)~\hbar$. 
From the intensities of the ND yrast transitions in our SD gated spectrum, the average entry point into the ND yrast region is found to be 2.24(15)~MeV and 14.7(4)~$\hbar$. The yrast transitions are taken from the ND level scheme of Ref.\cite{hubel}.
The energy and spin of the level fed by the 351~keV SD transition is, therefore, determined to be E$_{exit}$=5.7(5)~MeV and I$_{exit}$=17.8(8)~$\hbar$. 
Contributions to the uncertainty come from the calculated feeding statistical spectrum, the normalization to multiplicity,
uncertainty in the spin removed by the quasi-continuum decay-out component and the 
multipolarities of the unknown lines. 
The errors are added in quadrature.

\begin{figure}[htb]
  \centering
\vspace*{2.0cm}
\includegraphics[angle=-90,width=8.5 in]
	{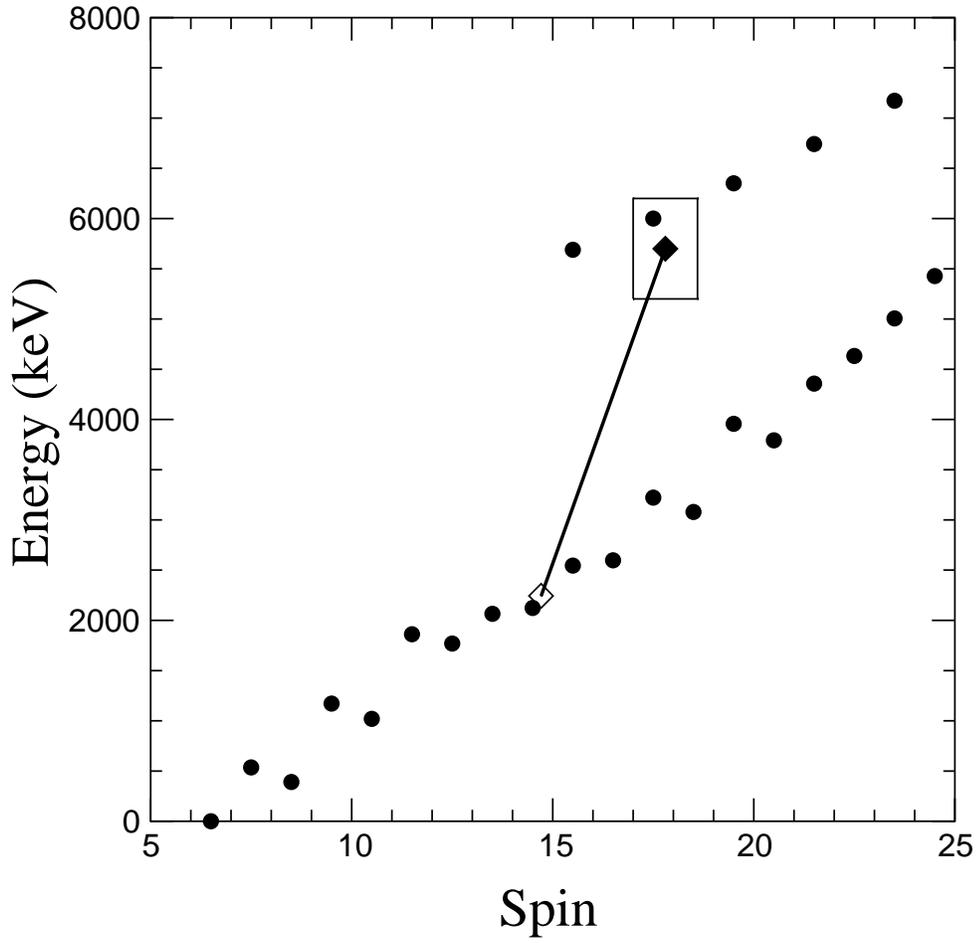}
\vspace*{1.0cm}
\caption{The spin and excitation energy of the level
fed by the 351-keV SD transition, obtained from the quasi-continuum analysis
(filled diamond); the box represents the uncertainty in spin and energy. The circles represent 
results from the one-step lines (see Fig.~\ref{fig:level}). The results from the two methods are 
in agreement. 
}
\label{fig:test}
\end{figure}

In Fig.~\ref{fig:test}, the experimental results from the 1-step 
linking transitions and from the quasi-continuum analysis are presented in a spin-energy diagram. The 
filled circles represent the yrast ND and SD levels, as given by the level scheme (Fig.~\ref{fig:level}), based on the one-step decay paths. The filled diamond denotes the SD level, which is fed by the 351 keV transition, and the open diamond the average 
entry point into the ND yrast band, as obtained from the quasi-continuum analysis. The box around the filled diamond shows the uncertainty in spin and energy from the quasi-continuum analysis. Clearly, 
the results from the 1-step linking transitions and from the quasi-continuum analysis are in good agreement with each other. This gives confidence about the spin and energy assignments. 
 
\section{Moments of inertia and spins}

\begin{figure}[htb]
  \centering
\vspace*{2.0cm}
\includegraphics[angle=-0,width=7.in]
              {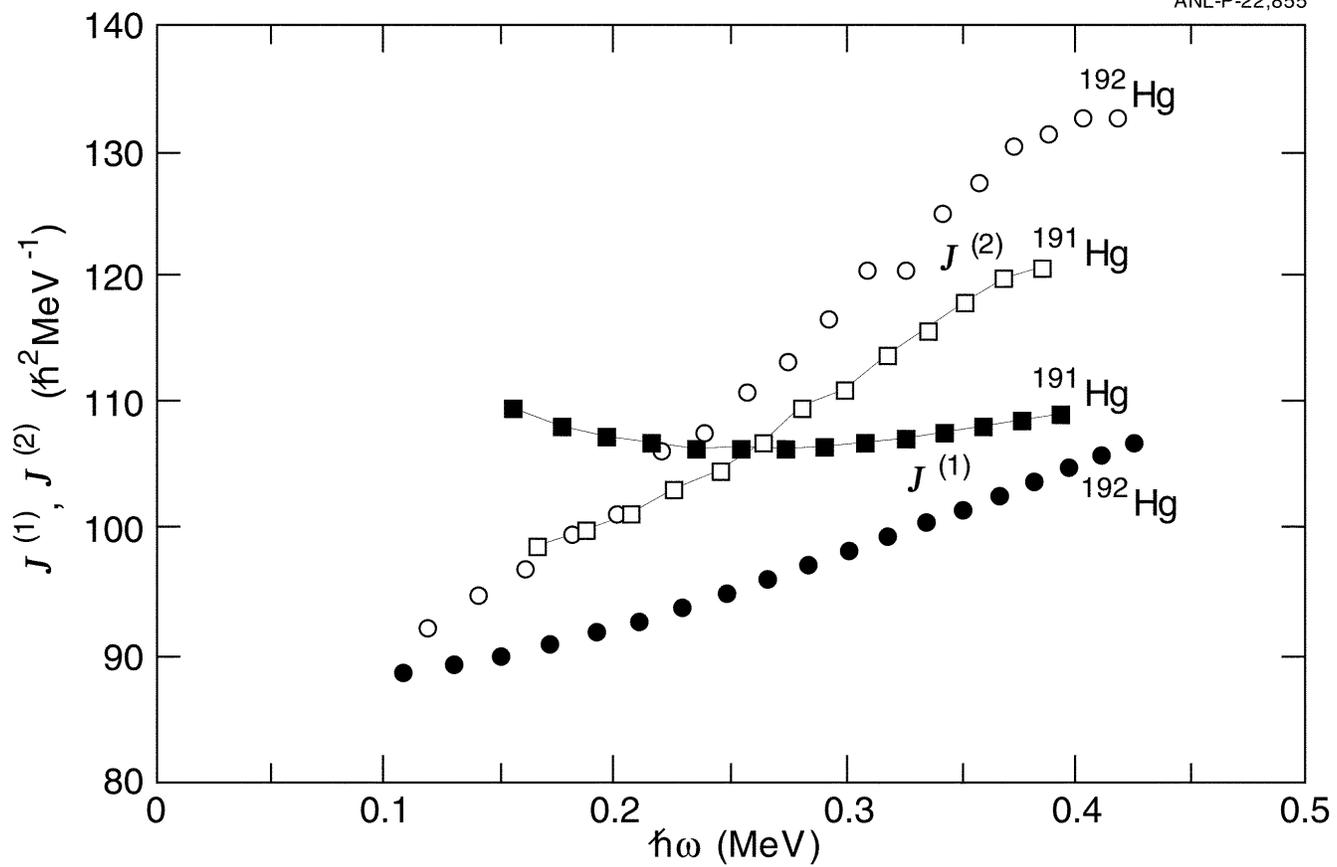}
\vspace*{1.0cm}
\caption{The dynamical and kinematic moments of inertia, $J^{(2)}$ (open symbols) and $J^{(1)}$ (filled symbols), for the yrast SD bands of $^{191}$Hg (squares) and $^{192}$Hg (circles).}
\label{j1j2}
\end{figure}

With the spins now known, the kinematic moment of inertia $J^{(1)}$ can also be determined.  Figure~\ref{j1j2} shows both $J^{(1)}$ and $J^{(2)}$ moments as a function of $\omega$. The dynamical moment of 
inertia $J^{(2)}$ can be expressed in terms of the Harris expansion\cite{harris}:
\begin{equation}
J^{(2)} = J_0 + 3J_1 \omega^2 = dI_x/d\omega    \label{Khoo1}
\end{equation}
Integration of Eq.~(\ref{Khoo1}) gives
\begin{equation}
I_x  = J_0\omega + J_1\omega^3 + i  \label{Khoo2}
\end{equation}
and
\begin{equation}                                                    
J^{(1)} = I_x/\omega = J_0 + J_1\omega^2 + i/\omega. \label{Khoo3}  
\end{equation}
Here $\omega$ is the rotational frequency, given by $\delta E/\delta I_x \approx E_{\gamma}/2$; $I_x$ is the spin perpendicular to the symmetry axis and $i$, the integration constant, represents the quasi-particle alignment. 
For $^{191}$Hg, fits of $J^{(2)}$ and $I_x$ \emph{vs.} $\omega$ with 
Eqs.~(\ref{Khoo1}) and (\ref{Khoo2}) give $J_0 = 92.6\hbar^2 MeV^{-1}, J_1 = 68.1\hbar^4MeV^{-3}, i = 2.4$. 
The behavior of $J^{(1)}$ of $^{191}$Hg is different from that of other A$\approx$190 SD bands, where the moments of inertia increase monotonically with $\omega$. The U-shaped curve of $J^{(1)}$ arises from  the $i$/$\omega$ term  (which causes the unusual rise at low $\omega$) and provides a characteristic signature for a band exhibiting finite alignment.
 Knowledge of the spins allows configuration assignments to be made on a solid foundation.  In the past, the assignments were largely based on the variation with rotational frequency $\omega$ of the dynamical moment of inertia $J^{(2)}$.  In only a handful of cases, spins were extracted using a fit method      
[using Eqs.~(\ref{Khoo1} and \ref{Khoo2})], with the assumption that $i$ =0]. 
For SD band 1 in $^{191}$Hg, which exhibits particle alignment, this method cannot be used and spins were proposed by Carpenter $\emph{et al.}$~\cite{carp95}, based on the entry spin into the ND yrast line after decay from the SD band.  The present work firmly establishes the spins and confirms the assignment of Ref.~\cite{carp95},            
thus validating the interpretation that SD band 1 in $^{191}$Hg is based on the N=7 $\em{j_{15/2}}$ [761]3/2 configuration. The alignment, $i = 2.4$, is an important ingredient in this assignment.
Together, this work and  Ref.~\cite{carp95} add confidence about the single-particle orbitals calculated by theory.  The Woods-Saxon potential gives this orbital as the yrast configuration at large deformation and at high spin. The neutron quasi-particle Routhians for $^{192}$Hg~\cite{Heenen-Janssens} suggest that this feature would also be given by HFB theory.                                                                            
However, details are not always correctly predicted, e.g., at the lowest  frequencies, there is a discrepancy of 2 $\hbar$ in alignment between  experiment and HFB theory for SD band 1 in $^{191}$Hg~\cite{Heenen-Janssens}.                                                                                                                   
Altogether, mean-field theories provide good descriptions of the general features   of SD bands in the mass 190 region, such as the rise of $J^{(2)}$ with $\omega$ (due partly to the N=7 orbital) and the convergence at high $\omega$ for most  nuclei.   This has been summarized in work by Fallon \emph{et al.}~\cite{Fallon}, which distills the main physics results.  

The assigned spin of the band is consistent with a favored $\alpha = - \frac{1}{2}$, $\em{j_{15/2}}$ particle configuration assignment, which is calculated~\cite{carp95} to be yrast at high spin. The experimental data do not allow for a parity assignment. However the $\em{j_{15/2}}$ assignment requires the SD band to have negative parity, 
implying M1 multipolarity for both of the one-step transitions. From neutron-capture data, it is known that, at E$_{\gamma}$~$\approx$~8 MeV, the decay is dominated by E1 transitions\cite{boll70}. However, in $^{191}$Hg the one-step transitions have significantly lower energy, E$_{\gamma}$~$\approx$~3 MeV. In fact, M1 transitions with similar energy have been observed to compete with E1 transitions in the decay out of the SD band in $^{194}$Pb\cite{lope94,boss97}.
In addition, in neutron capture experiments on $^{162}$Dy targets, the M1 strength was found to be comparable to the E1 strength at $E_{\gamma} \approx$ 3~MeV~\cite{becvar}. The scissor mode~\cite{Richter} probably accounts for the enhanced M1 strength.

\section{PROTON AND NEUTRON PAIRING GAPS}

The $\gamma$ spectrum of Fig.~\ref{fig:doyr} shows a region with depleted yield between 2.3 and 3.3~MeV.  Following D{\o}ssing \emph{et al.}~\cite{doss95}, this depletion can be explained by the reduction in level density in the interval 
from the ND yrast line up to the energy required to break the first pair of neutrons or protons. 
In Ref.~\cite{doss95} it is seen that the width of the depleted region in 
the $\gamma$ spectrum is around 1.5 times the average pairing gap.
The depleted region (which is most clearly defined 
by the decay-out transitions with $\Delta I = 1\hbar$)
occurs between 2.3 and 3.3 MeV, implying  
a pair gap of about 0.7 MeV. For the non-rotating nucleus, 
$\Delta_p$ (or $\Delta_n$) is approximately given by the five-point mass formula $\Delta_p^{(5)}$ for
a sequence of isotones (or isotopes)-- see, for example, Eq. (7) in Ref. \cite{duguet02}.
Around $^{191}$Hg, $\Delta_p\sim~0.9$ MeV (if one neglects mean-field 
contributions to $\Delta_p^{(5)}$, which are discussed in 
Ref.~\cite{duguet02}).
Although the information from the tail of the decay-out gamma-spectrum is quite uncertain, it
yields a pairing gap similar to that given by $\Delta_p^{(5)}$.

In Table~\ref{tab:theory}, the experimental SD excitation energies are presented given 
for $^{191,192}$Hg.
The excitation energies of 
the SD levels of $^{192}$Hg are given by two tentative decay-out 
pathways \cite{khoo99} from the 10$^+$ level, 
combined with limits imposed by the quasi-continuum analysis~\protect\cite{torb99},
giving $E^{10^{+}}_{SD} = 6.0^{+0.3}_{-0}$~MeV.
The SD bands are extrapolated to spin 2.9 and 0~$\hbar$, where the rotational 
frequencies are zero. (For $^{191}$Hg, Eq.~(\ref{Khoo2}) gives $I_x = i = 2.4$ at $\omega$~=~0, and  
$I= I_x + 1/2 = 2.9$.)
Table~\ref{tab:theory} also presents the SD excitation energies from theoretical 
calculations based on the self-consistent Hartree-Fock-Bogoliubov (HFB)  
approach with the effective Skyrme interactions, 
SkP, SLy4 and SkM$^*$, and the density-dependent zero-range interactions 
of Ref.~\cite{heen98} for the particle-particle (pairing) channel.
The theoretical results were 
extrapolated in the same way as the experimental values in  the case of $^{191}$Hg, 
while the theoretical value for the SD state in $^{192}$Hg was calculated directly 
for the ground state spin of 0~$\hbar$\cite{heen98}.
The calculations with the SLy4 interaction show the best overall agreement with the experimental data.

\begin{figure}[hbt]
  \centering
\vspace*{2.0cm}
  \includegraphics[angle=-90,width=8.5 in]
             {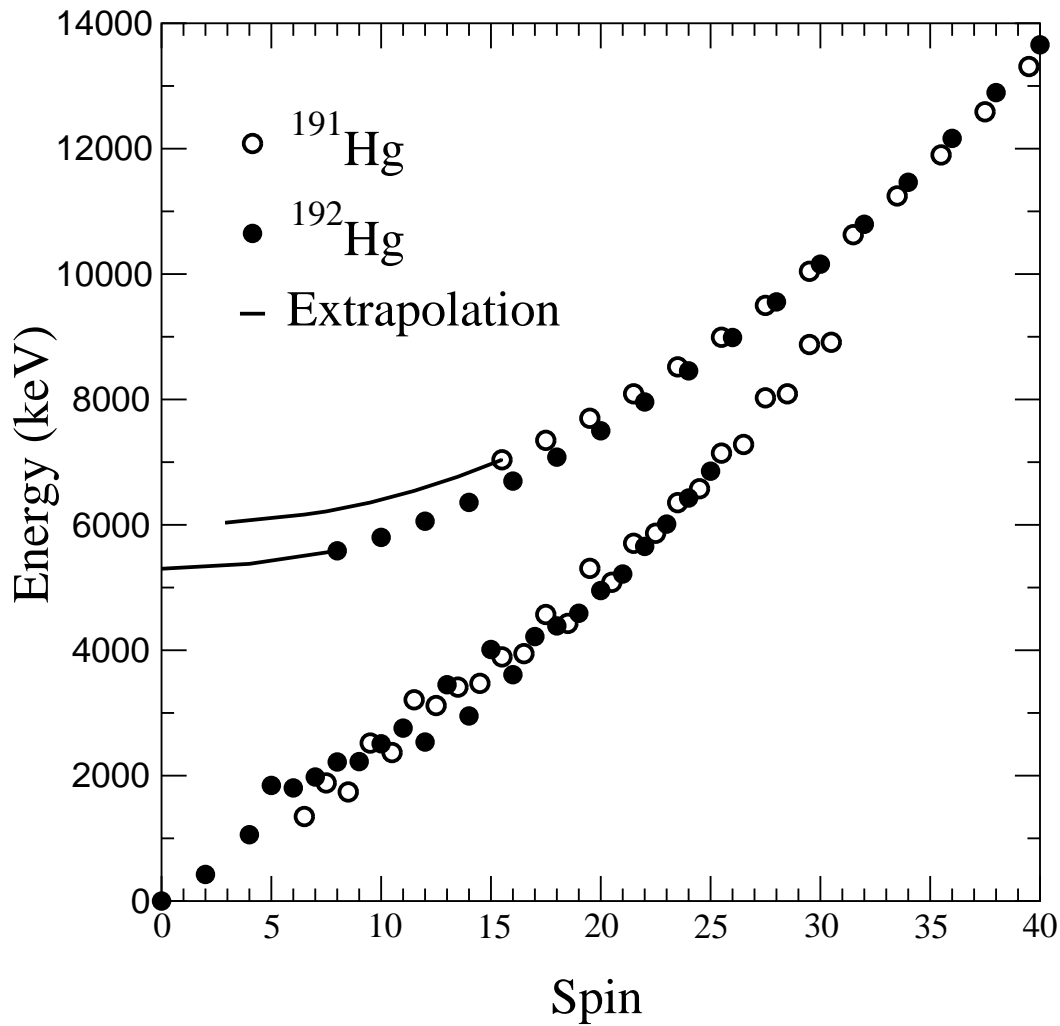}
\vspace*{1.0cm}
\caption{
The spins and excitation energies of the SD and ND yrast bands for $^{192}$Hg 
(filled circles) and $^{191}$Hg (open circles), plotted after correcting for the 
difference in the mass excess, 
i.e.  E($^{191}$Hg) = $E_x$($^{191}$Hg) - mass excess($^{192}$Hg) + mass excess ($^{191}$Hg).
The ground state of $^{192}$Hg is set at zero. The solid lines are the extrapolations of 
the $^{191,192}$Hg SD bands to spin 2.9 and 0~$\hbar$, where the rotational frequencies are zero.
The excitation energies of 
the SD levels of $^{192}$Hg are given by tentative decay-out pathways \protect\cite{khoo99}, 
combined with limits imposed by the quasi-continuum analyis~\protect\cite{torb99},
giving an uncertainty of $^{+0.3}_{-0}$~MeV.
}
\label{fig:mass}
\end{figure}

The excitation energies of SD states in odd-A and even-even nuclei give information on pairing in the SD well. 
In even-even nuclei pairing is stronger so the ground state has lower energy than that of the neighboring odd-even nuclei.  This can be seen by comparing the ND and SD yrast bands of $^{191}$Hg and $^{192}$Hg after accounting for the difference in the mass excess of the two nuclei - see Fig.~\ref{fig:mass}.
The ground state of $^{192}$Hg is taken as a reference, i.e. set to zero. 
One sees that the 0$^+$ state in $^{192}$Hg has a more negative mass excess than the
13/2$^+$ state in $^{191}$Hg, implying an extra binding of the even-even nucleus by $\sim$1.5~MeV.
The observed SD states are also shown in Fig.~\ref{fig:mass}.  
Here the energy of the SD ``ground'' state in $^{192}$Hg is lower than that in $^{191}$Hg by 0.8~MeV.
This smaller value (compared to 1.5~MeV for ND states), is consistent with reduced pairing in the SD well, as suggested before, e.g.~Ref.~\cite{carp95}, from the increase of the $J^{(2)}$ moment of inertia with frequency. However, in addition to pairing, mean-field effects (e.g. a change in the Fermi energy and a polarization energy) contribute to the binding energy 
\cite{duguet02}.  The convergence of the yrast lines of $^{191,192}$Hg around spin 10 and 25 $\hbar$, for ND and SD states, respectively, may be attributed to a reduction of pairing due to rotation.

\begin{figure}[htb]
\centering
\includegraphics[angle=-90,width=5.in]
               {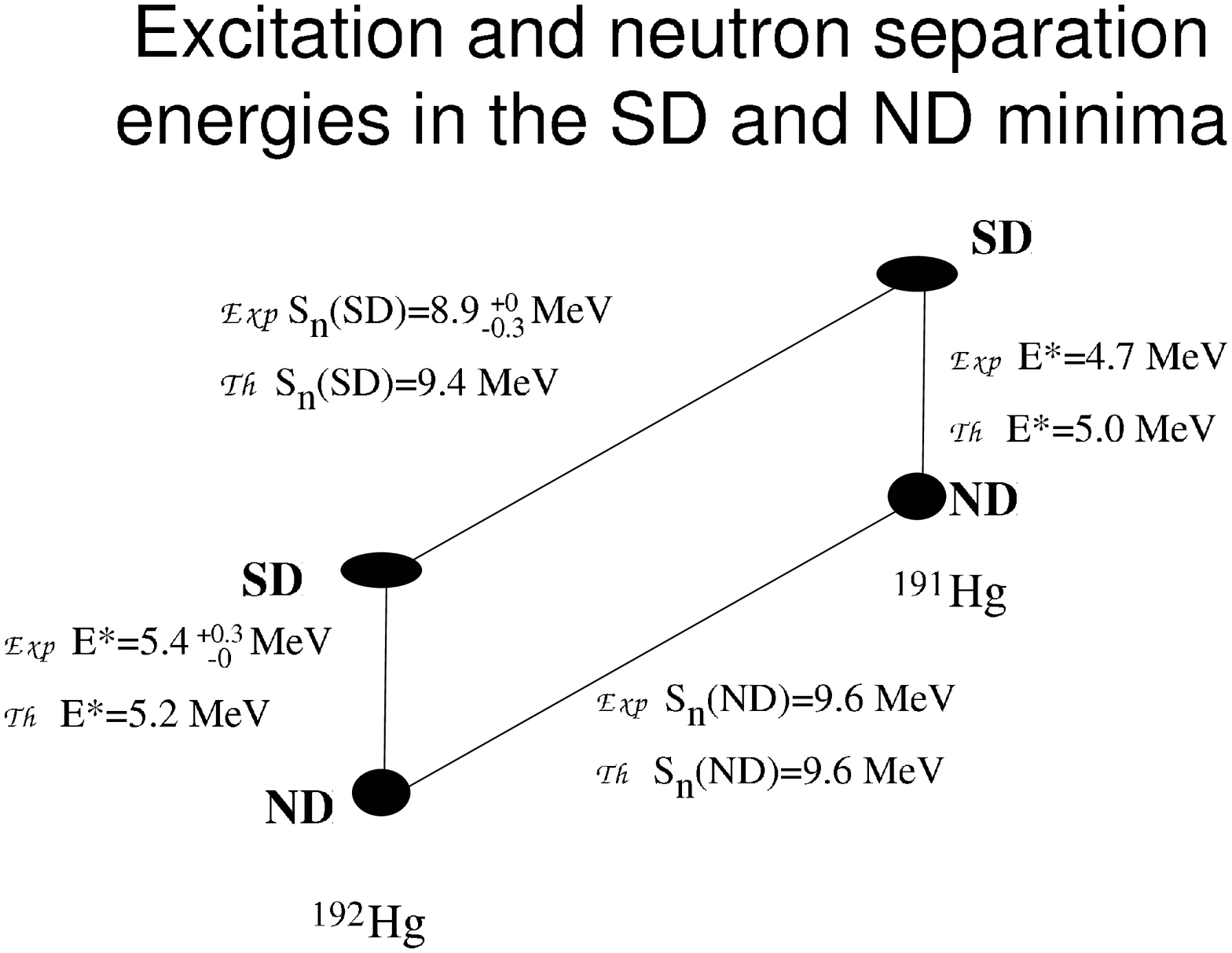}
\vspace*{1.0cm}
\caption{The experimental (top) and theoretical (bottom) values for the one neutron separation energy 
S$_n$(ND), S$_n$(SD) in the ND and the SD minima,
and the excitation energies E$^*$ of the SD bands in $^{191}$Hg and $^{192}$Hg. 
S$_n$(ND) values to the 13/2$^+$ level in $^{191}$Hg are given here. 
All values are taken at zero rotation, i.e. I = 0 $\hbar$ for both ND and SD in $^{192}$Hg and I = 13/2 $\hbar$ for ND and 2.9 $\hbar$ for SD in $^{191}$Hg. The SD bands had to be extrapolated to these spins (see text) - except for the theoretical value of $^{192}$Hg which was calculated directly for I = 0 $\hbar$\protect\cite{heen98}. The theoretical values are from HFB calculations with the SLy$_4$ Skyrme interaction.}
\label{fig:1nsep}
\end{figure}

An alternative but equivalent way to present the differences in binding
energies in Fig.~\ref{fig:mass} is in terms of the neutron separation 
energies $S_n$ in the ND and SD wells.  
[Note that $S_n$  =   mass excess ($^{191}$Hg) - mass excess($^{192}$Hg) + mass excess(neutron)].
Table~\ref{tab:theory} compares the experimental and theoretical 
neutron separation energies S$_n$ in the SD and ND wells. 
The S$_n$ values and SD excitation energies from HFB calculations with the 
SLy4 force are compared with experimental results in Figure~\ref{fig:1nsep}. 
The experimental neutron separation energy in the SD well is found to be 
S$_{n} = 8.9_{-0.3}^{+0}$~MeV, 
compared to S$_{n} = 9.6$~MeV in the ND well (to the 13/2$^+$ state). 
The difference of 0.7-1.0~MeV means 
that it is easier to remove a neutron from the SD well 
than from the ND well. 

As discussed above, part of the reduction in S$_n$ in the SD well is due to a
decrease in pairing, but other effects contribute as well.
In order 
to gauge the reliability
of extracting the pairing gap $\Delta^{pair}$ (and the difference in the ND and SD wells)
from nuclear masses, we write the equations for 
2-, 3- and 5-point
mass differences in the form discussed by Duguet $\emph{et al.}$ \cite{duguet02,duguet_np}. 

\begin{eqnarray}
\Delta^{(2)}(N) & = & (-1)^NS_n \nonumber\\ 
& \sim & \Delta^{pair}(N) + E^{pol}(N) + (-1)^{N+1}\lambda(N) \nonumber\\
\Delta^{(3)}(N) & \sim & \Delta^{pair}(N) + E^{pol}(N) + \frac{(-1)^N}{2} 
\frac{\partial \lambda(N)}{\partial N} \nonumber\\
\Delta^{(5)}(N) & \sim & \Delta^{pair}(N) + E^{pol}(N). \nonumber
\end{eqnarray}
Here N is the neutron number, $\Delta^{pair}$ the pairing gap, $E^{pol}$ the polarization energy 
due to time-reversal symmetry breaking (from the blocking of a single-particle level) and 
$\lambda$ the Fermi level. These equations show that, in order to deduce $\Delta^{pair}$
from experimental masses, theoretical values for each well are also required for 
$E^{pol}$, as well as for 
$\lambda$ or $\frac{\partial \lambda(N)}{\partial N}$ 
if $\Delta^{(2)}(N)$ or
$\Delta^{(3)}(N)$ are employed. It would
be best to use $\Delta^{(5)}(N)$, since that requires a calculation of only $E^{pol}(N)$. 
(The value of $E^{pol}(N)$ is around $\pm$100 keV, but there is some uncertainty in its 
calculation~\cite{duguet02a,satula98}.)
However, $\Delta^{(5)}(N)$ for SD states would need the SD excitation energies in 
5 consecutive nuclides, $^{190-194}$Hg, and would require new experimental SD energies 
in $^{190}$Hg 
(work on which is in progress\cite{siem04}), as well as in $^{193}$Hg.
For the ND \emph{ground states}, which have measured masses, $\Delta^{(5)}$ yields
\begin{eqnarray}
\Delta^{pair}_{ND} + E^{pol}_{ND}~\sim~1.16~MeV. \nonumber
\end{eqnarray}
Excellent agreement of 1.1 MeV is obtained using the experimental $\Delta^{(2)}$,
together with a theoretical $\lambda^{ND}_{HFB}$ = -8.4 MeV (obtained with the SLy4 force).
This agreement provides 
validation of $\lambda^{ND}_{HFB}$ for Hg nuclides around $^{192}$Hg, and is 
consistent with the reproduction of $S_{2n}$ values 
(within 0.2 MeV) for the ND ground states of nuclides~\cite{heen98} in this region with the SLy4 force.
For the SD well, $\Delta^{(2)}$ and $\lambda^{SD}_{HFB}$ ($-7.9$ MeV) give
\begin{eqnarray}
\Delta^{pair}_{SD} + E^{pol}_{SD} = 1.0^{+0}_{-0.3}~MeV, \nonumber
\end{eqnarray}
where the errors do not include the uncertainty in $\lambda^{SD}_{HFB}$.  This value of 
$\Delta^{pair}_{SD} + E^{pol}_{SD}$ is a direct indicator of pair correlations in the SD well.
There appears to a reduction of this value with respect to that in the ND well, but the 
uncertainties do not allow for a definitive conclusion.
\section*{Conclusions}
The spins and excitation energies of the yrast SD band in $^{191}$Hg have been determined from two single-step linking transitions and from the quasi-continuum spectrum that connects the SD and ND states. The results from the two methods are in good agreement, within the error bars, 
providing confidence about the spins and excitation energies of the yrast SD band. 
The SD level fed by the 351~keV SD transition has E$_{x}$~=~6000 keV and I~=~35/2 $\hbar$. Excitation energies and spins provide a stringent test of orbital assignments. The spin is consistent with that expected for a $\em{j_{15/2}}$ orbital configuration, previously assigned to this SD band\cite{carp95}. 

This is the first time that the excitation energies and spins have been determined for a SD band in an odd-even nucleus in the mass A~=~190 region. By comparing the results with those of neighboring even-even Hg nuclei, we have obtained information on pairing in the SD states.

 The neutron separation energies in the ND and SD wells have been extracted
by using data from $^{191,192}$Hg. 
The separation energy in the SD well is 0.7-1.0~MeV smaller than in the ND well,
due partly to a reduction in the pair gap $\Delta^{pair}$ with deformation and partly to an change in
$\lambda$.

We have compared the results with those from calculations based on Hartree-Fock-Bogoliubov (HFB) theory with different Skyrme interactions\cite{heen98}, and have found that the SLy4 interaction, which yields 
6.32~MeV for the excitation energy of the I~=~35/2 $\hbar$ SD level, gives the best agreement.  Similarly, the same interaction gives the best reproduction of the neutron separation energies in the ND and SD wells.
                          
\subsection*{Acknowledgments}
Discussions with I.~Ragnarsson and A.~Afanasjev are gratefully acknowledged. 
This research is supported in part by the U.S. Dept. of Energy under contract Nos. W-31-109-ENG-38 and 
DE-AC03-76SF00098. S. Siem acknowledges a NATO grant through the Research Council of Norway.\\

\end{multicols}

\newpage

\begin{table}[htb]
\centering
\begin{tabular}{|l|c|c|c|c|c|} 
component      & $<M>$ &$<E_{\gamma}>$&$<\delta I>$& $\Delta$ I& $\Delta$E \\
in $^{191}$Hg  &       &  MeV & $\hbar$  & $\hbar$     & MeV  \\\hline
Statisticals   & 2.18  & 1.87 & 0.66  & 1.44$\pm$0.14  & 4.06$\pm$0.4 \\ 
Quadrupoles    &  3.98 & 0.67 & 2.0   & 7.96$\pm$0.4   & 2.66$\pm$0.13 \\ 
M1/E2 dipoles  &  2.51 & 0.48 & 1.0   & 2.51$\pm$0.13  & 1.22$\pm$0.07  \\ 
SD transitions &       &      &       & 14.50$\pm$0.7  & 3.64$\pm$0.18 \\ \hline
Decay out:&&&&&\\
 quasi-continuum&1.95 &1.41  &  0.5  & 0.97$\pm$0.4   & 2.74$\pm$0.14 \\ 
non-yrast trans.&     &      &       & 2.0$\pm$0.4    & 0.70$\pm$0.05\\
ND yrast trans. &     &      &       & 8.21$\pm$0.4   & 2.24$\pm$0.14 \\ 
Decay-out point &     &      &       & 17.7$\pm$0.8   & 5.7 $\pm$0.5 \\ 
Level fed by 351keV trans. &&&       & 17.8$\pm$0.8   & 5.7$\pm$ 0.5  \\ 
Entry point     &     &      &       & 44.1$\pm$2.2   & 17.3 $\pm$0.86\\ 
\end{tabular}
\caption{The different feeding and decay components of the spectrum in coincidence with the yrast SD band in $^{191}$Hg. The total spin and energy removed, on average, by the different components are
 $\Delta$I and $\Delta$E; 
$\langle E_{\gamma}\rangle$ and $\langle \delta I_{\gamma}\rangle$
are, respectively, the average energy and 
spin removed per photon. From Monte-Carlo simulations, 
the quasi-continuum decay-out and statistical feeding components have, respectively, $< \delta$I$>$~=~0.5 and ~0.66 $\hbar$ per $\gamma$ ray. The unknown lines are defined as non-yrast discrete transitions.}
\label{tab:comp}
\end{table}

\begin{table}[htb]
\centering
\begin{tabular}{|l|c|c|c|c|c|}
       & $^{191}$Hg & $^{191}$Hg   &$^{192}$Hg& S$_{n}$ & S$_{n}$ \\
& E$^{*}$ (I=35/2) &E$^{*}$ (I=2.9)& E$^{*}$(I=0)& ND   & SD\\\hline
SkM$^*$      &  5.7   & 4.3$^a$       & 4.7     & 10.0$^b$    & 9.6  \\ 
SkP      &  5.5   & 4.2$^a$       & 4.6     & 9.1$^b$    & 8.6   \\ 
SLy4     &  6.3   & 5.0$^a$       & 5.2     & 9.6$^b$    & 9.4   \\ 
Exp.     &  6.0   & 4.7$^a$    & 5.4 $^{+0.3}_{-0}$& 9.6$^b$, 9.5$^c$  & 8.9 $_{-0.3}^{+0}$\\
\end{tabular}
\caption{The excitation energy E$^{*}$ of the yrast [761]3/2
SD levels in $^{191}$Hg (above the 13/2$^+$ state) from HFB calculations with different interactions 
and from experiment are given near the point of decay 
(I = 35/2 $\hbar$) and at I = 2.9 $\hbar$, where $\omega$~=~0 (marked by $^a$).
The excitation energy E$^{*}$ of the 
yrast SD band in $^{192}$Hg is given at its band head, I = 0 $\hbar$.
Extrapolations to the SD bandhead are described in the text.
The theoretical values for $^{192}$Hg are taken from Ref.~\protect\cite{heen98}. 
The theoretical and experimental neutron separation energies $S_n$ in the SD and ND wells are 
also given; 
$S_n$ values to the 13/2$^+$ level in 
$^{191}$Hg are indicated by $^b$ and to the 3/2$^{(-)}$ ground state by $^c$. 
The experimental masses for $^{192}$Hg and $^{191}$Hg are taken from Refs.~\protect\cite{rado99,auwa}, respectively.}
\label{tab:theory}
\end{table}


\newpage

\newpage

\newpage

\newpage

\newpage

\newpage

\newpage

\newpage

\newpage

\newpage
\vspace*{1.0cm}

\end{document}